Corticosteroid Activation of Atlantic Sea Lamprey Corticoid Receptor: Allosteric Regulation by the N-terminal Domain

Yoshinao Katsu[1,] *, Xiaozhi Lin[2], Ruigeng Ji[2], Ze Chen[2], Yui Kamisaka[2], Koto Bamba[1], Michael E. Baker[3,4,] *

[1] Faculty of Science

Hokkaido University

Sapporo, Japan

[2] Graduate School of Life Science

Hokkaido University

Sapporo, Japan

[3] Division of Nephrology-Hypertension

Department of Medicine, 0693

University of California, San Diego

9500 Gilman Drive

La Jolla, CA 92093-0693

Center for Academic Research and Training in Anthropogeny (CARTA) [4]

University of California, San Diego

La Jolla, CA 92093

*Correspondence to
Y. Katsu; E-mail: ykatsu@sci.hokudai.ac.jp
M. E. Baker; E-mail: mbaker@health.ucsd.edu

**Abstract**

Lampreys are jawless fish that evolved about 550 million years ago at the base of the vertebrate line. Modern lampreys contain a corticoid receptor (CR), the common ancestor of the glucocorticoid receptor (GR) and mineralocorticoid receptor (MR), which first appear in cartilaginous fish, such as sharks. Until recently, 344 amino acids at the amino terminus of adult lamprey CR were not present in the lamprey CR sequence in GenBank. A search of the recently sequenced lamprey germline genome identified two CR sequences, CR1 and CR2, containing the 344 previously un-identified amino acids at the amino terminus. CR1 also contains a novel four amino acid insertion in the DNA-binding domain (DBD). We studied corticosteroid activation of CR1 and CR2 and found their strongest response was to 11-deoxycorticosterone and 11-deoxycortisol, the two circulating corticosteroids in lamprey. Based on steroid specificity, both CRs are close to elephant shark MR and distant from elephant shark GR. HEK293 cells transfected with full-length CR1 or CR2 and the MMTV promoter have about 3-fold higher steroid-mediated activation compared to HEK293 cells transfected with these CRs and the TAT3 promoter. Deletion of the amino-terminal domain (NTD) of lamprey CR1 and CR2 to form truncated CRs decreased transcriptional activation by about 70% in HEK293 cells transfected with MMTV, but increased transcription by about 6-fold in cells transfected with TAT3, indicating that the promoter has an important effect on NTD regulation of CR transcription by corticosteroids.

**1. Introduction.**

The sea lamprey (Petromyzon marinus), belongs to an ancient group of jawless vertebrates known as cyclostomes, which last shared a common ancestor with vertebrates about 550 million years ago (1–3). As an outgroup to the jawed vertebrates, lampreys are important for studying early events in the evolution of vertebrates (1–8). Lampreys and hagfish, the other extant cyclostome lineage, contain a corticoid receptor (CR), which is the common ancestor to both the mineralocorticoid receptor (MR) and the glucocorticoid receptor (GR) (4,9). The MR and GR first appear as separate steroid receptors in sharks and chimeras (4,9–13). The CR, MR and GR belong to the nuclear receptor family of transcription factors, which also contains the progesterone receptor, estrogen receptor and androgen receptor (14–18).

Early studies from Thornton's laboratory provided important insights into corticosteroid activation of lamprey CR (4). These studies reported that several corticosteroids (Figure 1), including aldosterone, cortisol, corticosterone,11-deoxycorticosterone and 11-deoxycortisol activated lamprey CR. Although aldosterone, the physiological mineralocorticoid in humans and other terrestrial vertebrates (13,19–25), is the strongest activator of lamprey CR (4), neither lampreys nor hagfish synthesize aldosterone (4). Later studies with lamprey revealed that 11-deoxycortisol and 11-deoxycorticosterone (Figure 1) are the circulating corticosteroids in sea lamprey (26–28).

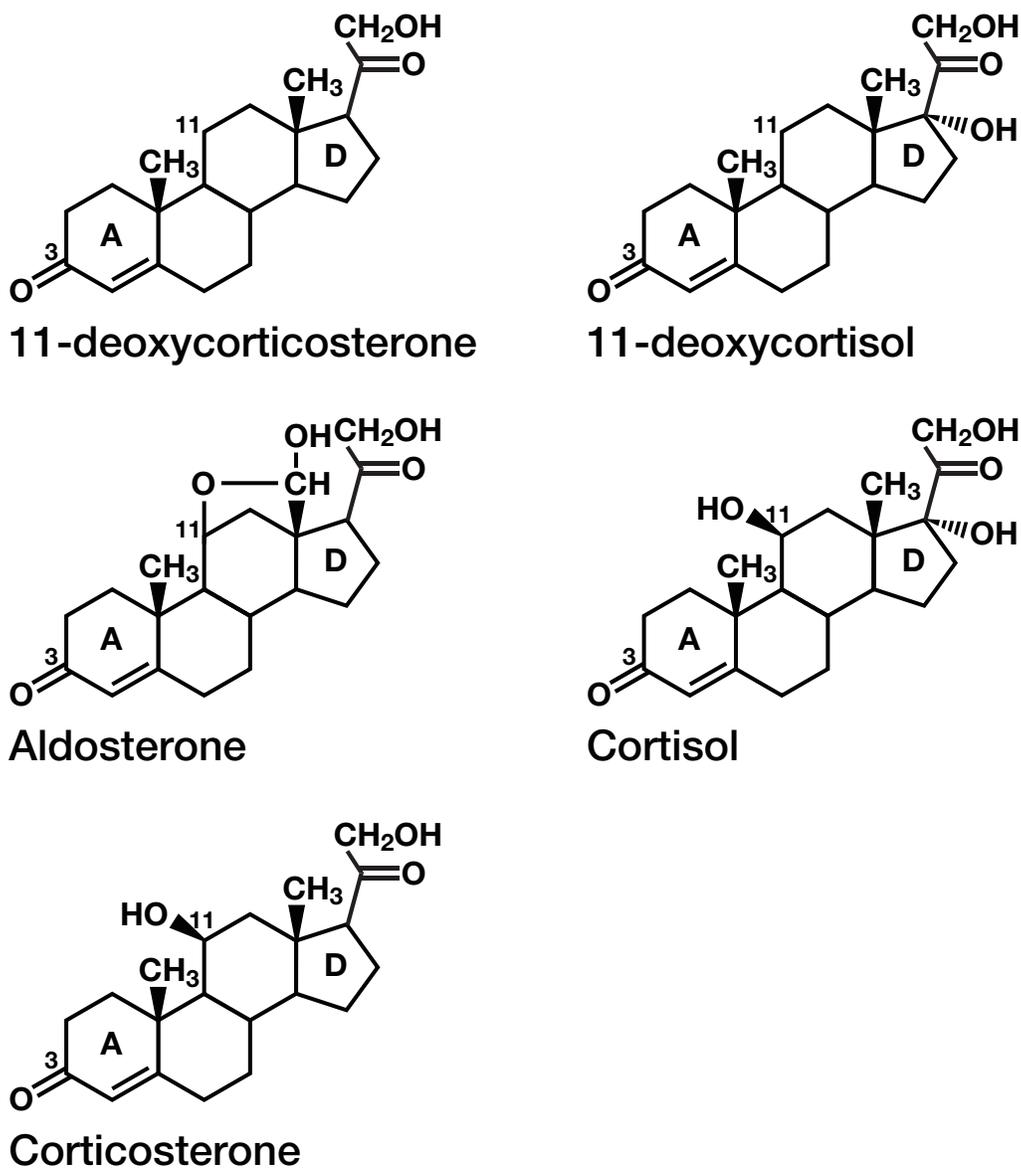

**Figure 1. Structures of Corticosteroids.** 11-deoxycorticosterone and 11-deoxycortisol are circulating corticosteroids in lamprey (26–28). Aldosterone is the physiological mineralocorticoid in terrestrial vertebrates (13,19–24). Aldosterone is not synthesized by lampreys (4). Cortisol and corticosterone are glucocorticoids in terrestrial vertebrates and fish (29–32).

Until recently, due to complexities in sequencing and assembly of the lamprey's highly repetitive and GC rich genome (3,33,34), DNA encoding 344 amino acids at the amino terminus of lamprey CR was present on a separate contiguous sequence that was not joined with the rest of the lamprey CR sequence and therefore not retrieved with BLAST searches of GenBank. The recent sequencing of the sea lamprey germline genome (35) provided contiguous DNA for two CR isoforms, CR1 and CR2, that encode the previously unassembled 344 amino acids at the amino terminus. The sequences of lamprey CR1 and CR2 reveal that like other nuclear receptors, lamprey CR is a multi-domain protein consisting of an N-terminal domain (NTD), a central DNA-binding domain (DBD), a hinge domain and a C-terminal ligand-binding domain (LBD) (4,14,36,37) (Figure 2). The DBD and LBD in lamprey CR are conserved in vertebrate MRs and GRs (Figure 2) (9,38), while their sequences in the NTD and hinge domains have diverged (Figure 2). The only sequence difference between lamprey CR1 and CR2 is a four amino acid insertion in the DBD in CR1 is not present in the DBD in CR2 or in other GRs and MRs.

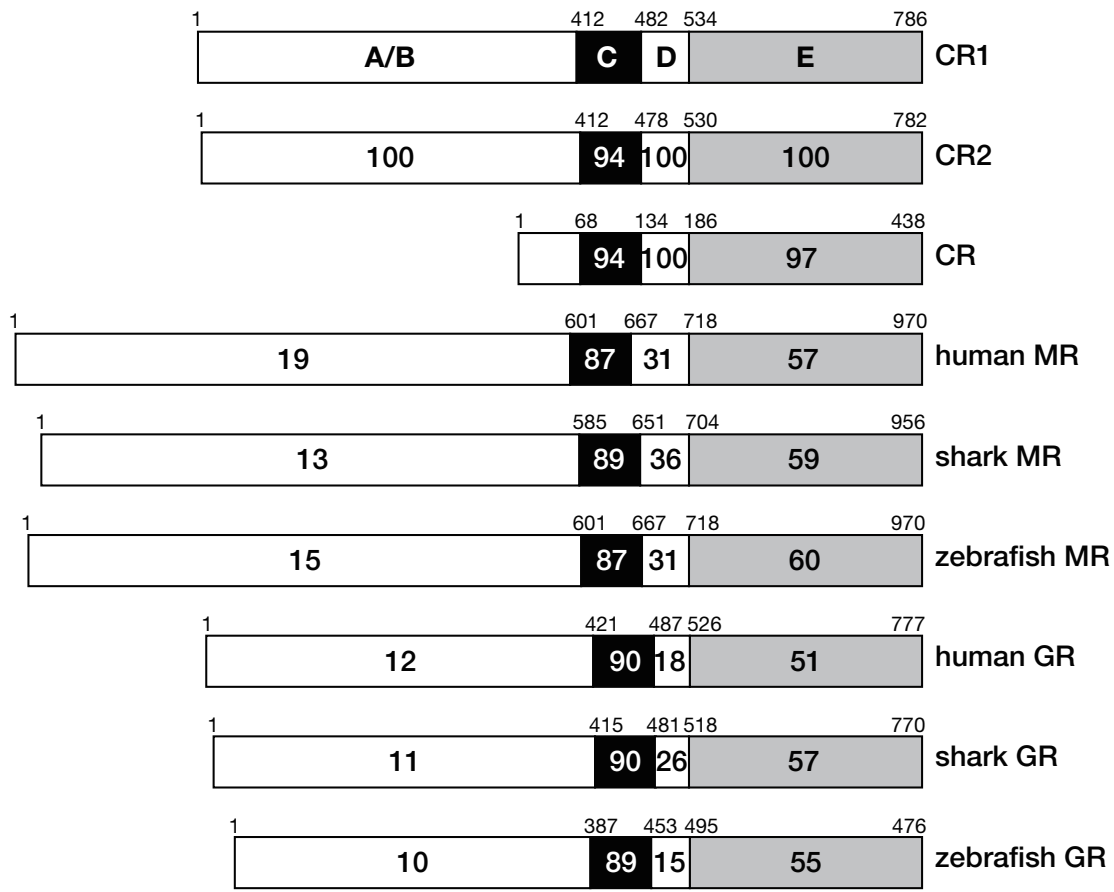

**Figure 2. Comparison of the functional domains of lamprey CR1 to CR2, previously cloned lamprey CR and selected vertebrate MRs (human, elephant shark, zebrafish) and GRs (human, elephant shark, zebrafish).** Lamprey CR1 and human MR and GR have 87% and 90% identity in the DBD and 57% and 51% identity in the LBD, respectively. Lamprey CR1 and elephant shark MR and GR have 89% and 90% identity in the DBD and 59% and 57% identity in the LBD, respectively. Lamprey CR1 and zebrafish MR and GR have 87% and 89% identity in the DBD and 60% and 55% identity in the LBD, respectively. This strong conservation of the DBD and LBD contrasts with the low sequence identity of 10-19%% and 15-36% between their NTD and hinge domains, respectively. Lamprey CR (4,38) is a partial CR sequence that was cloned from adult lamprey.

With the full sequences of the two lamprey CR isoforms in hand, we investigated four questions about this receptor that shares common ancestry with gnathostome GRs and MRs (9,13,38). First, what is the response of lamprey CR to a panel of physiological corticosteroids (aldosterone, cortisol, corticosterone, 11-deoxycorticosterone and 11-deoxycortisol) (Figure 1) for vertebrate GRs and MRs, and, second, how does this response compare to the response to these steroids by elephant shark GR and MR (12,39)? Comparison of corticosteroid activation of lamprey CR with elephant shark GR and MR can provide insights into the evolution of corticosteroid specificity in the GR and MR.

Third, what is the role, if any, of the NTD in transcriptional activation of lamprey CR? The NTD on human GR contains an activation function 1 (AF1) domain that is very important in GR activation by steroids (40–48). The NTD on human MR also contains an AF1 domain, although it is a much weaker activator of the MR (41,49–51) compared to the AF1 on human GR. The NTD on elephant shark GR also contains a strong AF1 (12). The low sequence identity (less than 20%) between the NTD in lamprey CR and in elephant shark MR and GR raises the question: Is there AF1 activity in lamprey CR or did a strong AF1 evolve after CR duplication and divergence to form vertebrate GR and MR?

Fourth, what is the role of the MMTV (52,53) and TAT3 (54) promoters in transcriptional activation of lamprey CR? That is, does lamprey CR have different responses to corticosteroids in cells co-transfected with MMTV or TAT3 promoters, as we found for cells co-transfected with either MMTV or TAT3 and either elephant shark MR (55) or GR (12). Comparison of activation of lamprey CR in cells with either MMTV or TAT3 with that of elephant shark GR and MR could indicate whether lamprey CR was closer to the GR or to the MR, and thus shed light on the evolution of the GR and MR from their common ancestor.

In this report, we used two metrics for evaluating activation by steroids of lamprey CR and other receptors. The first metric was the half maximal response (EC50) to various steroids, and the second metric was the strength (fold-activation) of transcription. Combined, these two metrics provide insights into the relevance of a steroid as a physiological ligand for the CR.

Our initial experiments focused on lamprey CR1 because RNA-Seq analysis indicates that CR1 is more highly expressed than CR2 (greater than 99%) in lamprey tissue. However, CR1 and CR2 have similar EC50s for corticosteroids. We find that the EC50s for activation by 11-deoxycorticosterone and 11-deoxycortisol, the two circulating corticosteroids in lampreys (26–28), of full-length lamprey CR1 in HEK293 cells with MMTV were 0.16 nM and 1.5 nM, respectively. These are the lowest EC50s for CR1 among the corticosteroids that we studied. Aldosterone, cortisol and corticosterone had EC50s from 2 nM to 9.9 nM for activation of CR1 in cells with MMTV. For truncated CR1, which lacks the NTD, the EC50 of 11-deoxycorticosterone for lamprey CR1 in HEK293 cells with MMTV was 0.4 nM, while EC50s of the other corticosteroids for lamprey CR1 increased from 3 to 6-fold.

Comparison of corticosteroid activation of CR with that of elephant shark MR and GR reveals that full-length and truncated elephant shark MR and the CR have similar EC50s for 11-deoxycorticosterone, 11-deoxycortisol and other corticosteroids, in contrast to full-length and truncated elephant shark GR, which has a negligible response to 11-deoxycortisol and weak responses to other corticosteroids (12), indicating that, based on steroid specificity, elephant shark MR is a closer to CR1 and CR2 than is elephant shark GR, which has diverged more from its common ancestor with the MR.

Interestingly, we found differences between the effect of the MMTV and TAT3 promoters on fold-activation of transcription of full-length CR and of truncated CR. Unexpectedly, fold-activation of full-length CR1 to corticosteroids was about 3 to 4-fold higher in cells with the MMTV promoter than in cells transfected with the TAT3 promoter. Removal of the NTD on CR1 decreased fold-activation by corticosteroids of truncated CR1 in cells with MMTV by about 70% indicating that there is an activation function in the NTD. In contrast, compared to full-length CR1, transcriptional activation by corticosteroids of truncated CR1 in cells with TAT3 increased by about 6-fold, indicating that the CR1 NTD represses steroid activation in the presence of TAT3. These data indicate that allosteric regulation by the NTD evolved before the evolution of the GR and MR from their common ancestor in cartilaginous fishes, with divergence of specificity for various corticosteroids evolving in elephant shark GR and MR (12).

**RESULTS**

**Comparison of functional domains on lamprey CR to domains on selected vertebrate MRs and GRs.**

To begin to understand the evolution of the MR and GR from the CR, we compared the sequences of functional domains on lamprey CRs with corresponding domains in human, elephant shark and zebrafish MRs and GRs (Figure 2). The sequences of LBD and hinge domain of lamprey CR1 and CR2 have more similarity to the LBD and hinge domain on vertebrate MRs than to the GRs as has been reported for lamprey CR (9,13,38,56). The strong conservation of the DBD and LBD contrasts with the low sequence identity of 10-19% between the NTDs and 15-36% between the hinge domains. The low similarity of the NTD on the CR to the NTD on elephant shark MR and GR of 13% and 12%, respectively, indicates that there was rapid evolution of the NTD during the divergence of a distinct MR and GR from their CR ancestor. Moreover, the NTD of elephant shark GR has only 21% sequence identity with the NTD on elephant shark MR, additional evidence for rapid evolution of the NTD early in the evolution of the MR and GR (12).

Although most of the sequence divergence among these receptors occurred in the NTD and hinge domain, there is an insertion of four amino acids in the DBD of CR1 that is not found in the DBD of either CR2 or the MR and GR (Figure 3). Otherwise, the DBD in both CRs is highly conserved in the MR and GR, with CR2 being closest to the DBD in the other MRs and GRs. Based on this analysis of the DBD, CR2 appears to be related to the common ancestor of the MR and GR.

```
lamprey CR1:        CLICSDEASGCHYGVLTCGSCKVFFKRAVEGTRQGQHNYLCAGRNDCIIDKIRRKNCPACRLRKCIQAGM
lamprey CR2:        CLICSDEASGCHYGVLTCGSCKVFFKRAVEG----QHNYLCAGRNDCIIDKIRRKNCPACRLRKCIQAGM
human MR:           CLVCGDEASGCHYGVVTCGSCKVFFKRAVEG----QHNYLCAGRNDCIIDKIRRKNCPACRLQKCLQAGM
elephant shark MR:  CLVCSDEASGCHYGVLTCGSCKVFFKRAVEG----QQNYLCAGRNDCIIDKIRRKNCPACRLRKCLKAGM
zebrafish MR:       CLVCGDEASGCHYGVVTCGSCKVFFKRAVEG----QHNYLCAGRNDCIIDKIRRKNCPACRVRKCLQAGM
human GR:           CLVCSDEASGCHYGVLTCGSCKVFFKRAVEG----QHNYLCAGRNDCIIDKIRRKNCPACRYRKCLQAGM
elephant shark GR:  CLVCSDEASGCHYGVLTCGSCKVFFKRAVEG----QHNYLCAGRNDCIIDKIRRKNCPACRFRKCLQAGM
zebrafish GR:       CLVCSDEASGCHYGVLTCGSCKVFFKRAVEG----QHNYLCAGRNDCIIDKIRRKNCPACRFRKCLMAGM
```

**Figure 3. Comparison of the DNA-binding domains on lamprey CR1, CR2, elephant shark MR and GR and human MR and GR.** The DNA-binding domain of lamprey CR1 has a unique insertion of four amino acids. Otherwise the DNA-binding domain on CR1 and CR2 are identical. Differences between the sequence of lamprey CR and selected vertebrate MRs and GRs are shown in red.

**RNA-Seq analysis indicates that lamprey CR1 is the predominate CR expressed in lamprey.**

To gain an insight into the relative biological importance of CR1 and CR2 in lamprey, we used RNA-Seq analysis to investigate the relative expression of lamprey CR1 and CR2 in lamprey tissues using databases in GenBank (3,35). As shown in Table 1, RNA-Seq analysis reveals that expression of lamprey CR1 is substantially higher than CR2. Indeed, CR1 expression is over 99% higher than CR2 in the intestine and kidney in the larval stage, in the parasitic and adult stages, as well as in the parasitic liver and in the adult intestine, kidney and brain.

Table 1. RNA-Seq Analysis of Expression of Lamprey CR1 and CR2.

|  | Larval Stage | | Parasitic Stage | | | | Adult Stage | | |
| --- | --- | --- | --- | --- | --- | --- | --- | --- | --- |
|  | Intestine | Kidney | Proximal intestine | Distal intestine | Kidney | Liver | Brain | Intestine | Kidney |
| CR1 | 21.78 | 10.36 | 23.23 | 19.44 | 21.49 | 24.96 | 2471 | 35.48 | 30.32 |
| CR2 | 0.1 | 0.13 | 0.22 | 0.32 | 0.09 | 0 | 0 | 0.2 | 0.35 |

Single-end RNA-Seq reads of sea lamprey, *Petromyzon marinus* for intestine and kidney from larval stage, intestine, kidney and liver from parasitic stage, brain, intestine and kidney from adult stage, were downloaded from database of National Center for Biotechnology Information (accession number: PRJNA50489). The relative measure of transcript abundance is FPKM (fragments per kilobase of transcript per million mapped reads) (57). FPKM values were estimated by normalizing gene length, followed by normalizing for sequencing depth.

**Corticosteroid-dependent and promoter-dependent activation of full-length and truncated lamprey CR1 and CR2.**

To gain a quantitative measure of corticosteroid activation of full-length and truncated lamprey CR1 and CR2, we determined the concentration dependence of transcriptional activation by corticosteroids of full-length lamprey CR1 transfected into HEK293 cells with either an MMTV-luciferase promoter (Figure 4A) or a TAT3 luciferase promoter (Figure 4B). A parallel study was done with truncated lamprey CR1 (Figure 4 C, D). Luciferase levels were used to calculate an EC50 value and fold-activation for each steroid for lamprey CR1 in HEK293 cells with either MMTV (Table 2) or TAT3 (Table 2).

Similar experiments were performed for corticosteroid activation of full-length and truncated lamprey CR2 in HEK293 cells, containing either the MMTV or TAT3 promoters (Figure 5). EC50 values and fold-activation for each steroid for lamprey CR2 in the presence of either MMTV or TAT3 are shown in Table 2.

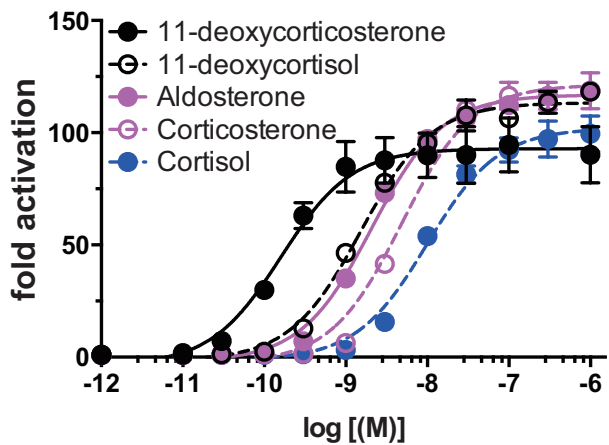
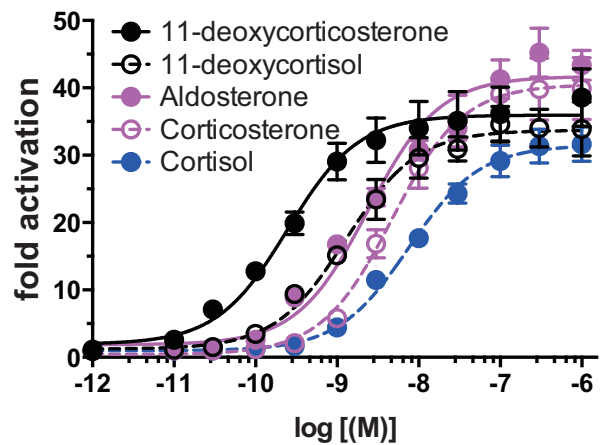
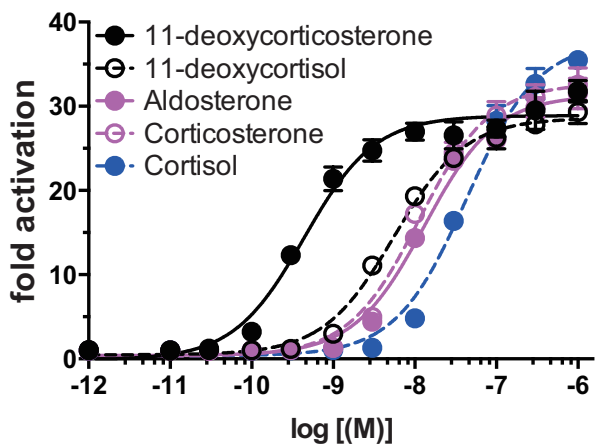
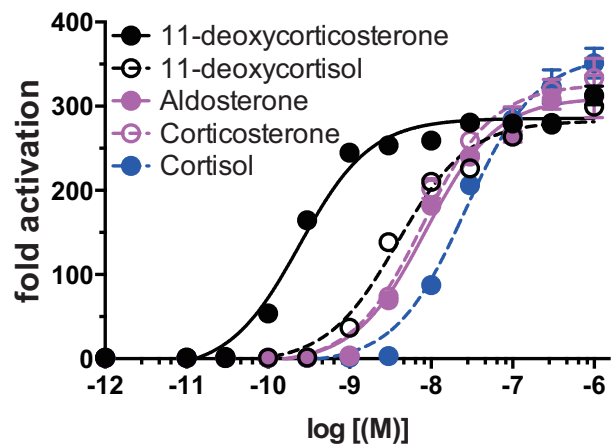

**Fig. 4. Concentration-dependent transcriptional activation by corticosteroids of full length and truncated lamprey CR1.** Plasmids for full-length or truncated lamprey CR1 were expressed in HEK293 cells with either an MMTV-luciferase promoter or a TAT3-luciferase promoter. Cells were treated with increasing concentrations of either aldosterone, cortisol, corticosterone, 11-deoxycortisol, 11-deoxycorticosterone or vehicle alone (DMSO). Results are expressed as means ± SEM, n=3. Y-axis indicates fold-activation compared to the activity of vector with vehicle (DMSO) alone as 1. A. Full-length CR1 with MMTV-luc. B. Full-length

CR1 with TAT3-luc.  C. Truncated lamprey CR1 with MMTV-luc.  D. Truncated lamprey CR1 with TAT3-luc.

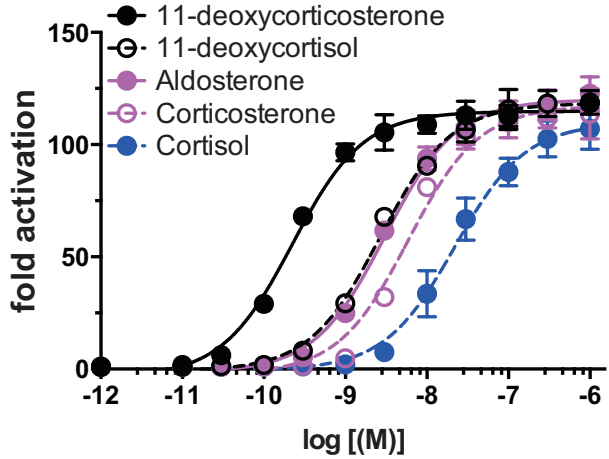
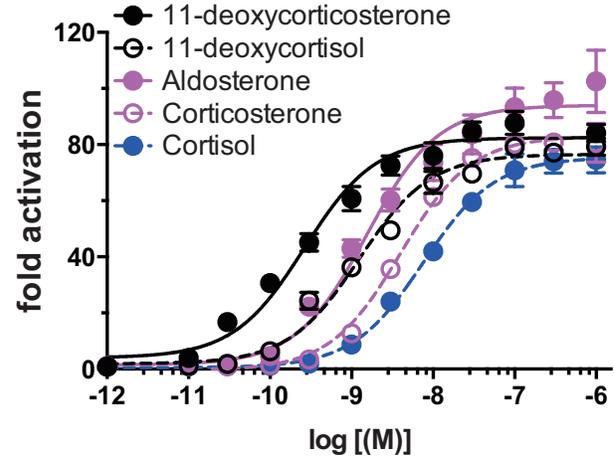
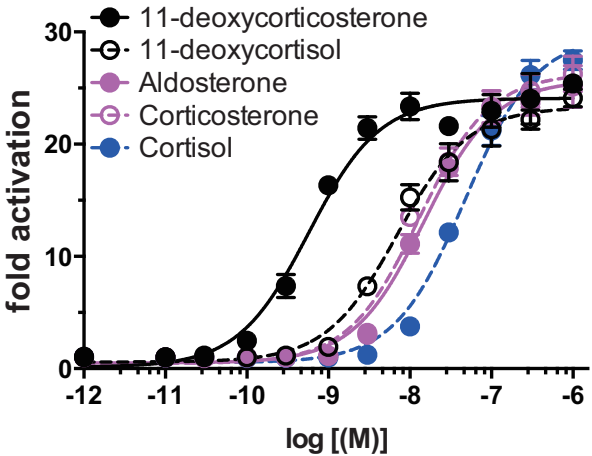
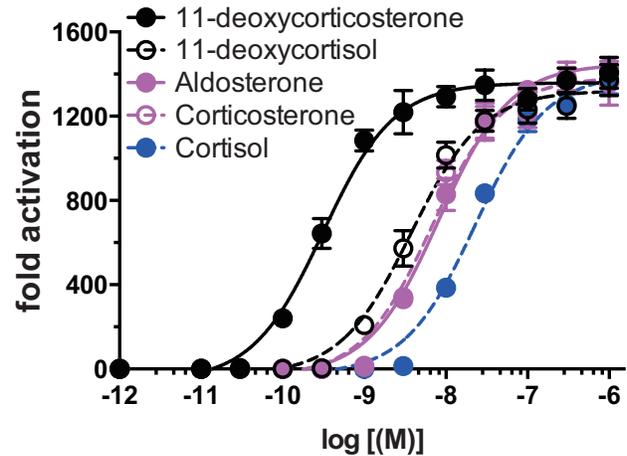

**Fig. 5. Concentration-dependent transcriptional activation by corticosteroids of full length and truncated lamprey CR2.** Plasmids for full-length or truncated lamprey CR2 were expressed in HEK293 cells with either an MMTV-luciferase promoter or a TAT3-luciferase promoter.  Cells were treated with increasing concentrations of either aldosterone, cortisol, corticosterone, 11-deoxycortisol, 11-deoxycorticosterone or vehicle alone (DMSO).  Results are expressed as means ± SEM, n=3.  Y-axis indicates fold-activation compared to the activity of vector with vehicle (DMSO) alone as 1.  A. Full-length CR2 with MMTV-luc.  B. Full-length

CR2 with TAT3-luc. C. Truncated lamprey CR2 with MMTV-luc. D. Truncated lamprey CR2 with TAT3-luc.

The results in Figures 4 and 5 and Table 2 show that transcriptional activation of full-length and truncated lamprey CR1 and CR2 in HEK293 cells is different in the presence of the MMTV and TAT3 promoters. In cells containing the MMTV promoter and full-length CR1, 11-deoxycortisol and 11-deoxycorticosterone, the two circulating corticosteroids in lamprey (26–28) have EC50s of 1.5 nM and 0.16 nM respectively (Table 2). These are the lowest EC50s of the tested corticosteroids, although aldosterone, corticosterone and cortisol also have low EC50s, which vary from 2.1 nM to 9.9 nM. In cells with MMTV, loss of the NTD raises the EC50 for corticosteroids, although the values for 11-deoxycorticosterone (0.4 nM) and 11-deoxycortisol (5.6 nM) are low. Cortisol has EC50 of 43 nM. Loss of the NTD results in a decline of fold-activation of CR1 by about 75%.

Analysis of Figure 4A and 4B and Table 2 shows that activation of full-length CR1 by corticosteroids is about 3-fold higher in HEK293 cells with the MMTV promoter than in cells with the TAT3 promoter. Removal of the NTD from CR1 leads to a decrease in fold-activation of about 75% by corticosteroids in cells with MMTV promoter (Figure 4C vs 4A). In contrast, in HEK293 cells with truncated CR1 and TAT3, corticosteroid stimulated transcription is about 7-fold higher than in cells with full-length CR1 and TAT3 (Table 2). Moreover, in cells with truncated CR1 and TAT3 corticosteroid-mediated transcriptional activation is about 50-fold higher than in cells with truncated CR1 and MMTV (Table 2).

**Table 2. Corticosteroid Activation of Lamprey CR1 and CR2 in HEK293 cells with an MMTV promoter or a TAT3 promoter.**

| MMTV-luc | | DOC | 11-deoxycortisol | ALDO | Corticosterone | Cortisol |
|---|---|---|---|---|---|---|
| Lamprey CR1 Full Seq. | EC50 (nM) | 0.16 | 1.5 | 2.1 | 4.8 | 9.9 |
| | Fold-Activation (±SEM) | 95 (± 12) | 106 (± 2.5) | 112 (± 2.5) | 116 (± 6.0) | 92 (± 5.3) |
| Lamprey CR1 Truncated Seq. | EC50 (nM) | 0.4 | 5.6 | 13.2 | 11.2 | 43.1 |
| | Fold-Activation (±SEM) | 27 (±1.0) | 26 (± 1.3) | 26 (± 0.9) | 29 (± 1.7) | 28 (± 1.8) |
| Lamprey CR2 Full Seq. | EC50 (nM) | 0.2 | 2.7 | 3.1 | 5.9 | 14.2 |
| | Fold-Activation (±SEM) | 113 (± 4.8) | 116 (± 9.0) | 113 (± 6.4) | 110 (± 7.0) | 88 (± 6.2) |
| Lamprey CR2 Truncated Seq. | EC50 (nM) | 0.6 | 7.0 | 14.9 | 13.0 | 47.8 |
| | Fold-Activation (±SEM) | 23.0 (±0.8) | 21.3 (±0.9) | 23.3 (± 0.8) | 23.3 (± 0.8) | 21.2 (± 0.8) |
| **TAT3-luc** | | **DOC** | **11-deoxycortisol** | **ALDO** | **Corticosterone** | **Cortisol** |
| Lamprey CR1 Full Seq. | EC50 (nM) | 0.24 | 1.3 | 2.4 | 4.7 | 7.5 |
| | Fold-Activation (±SEM) | 36 (± 4) | 35 (± 2.6) | 41 (± 2.9) | 39 (± 2.4) | 29 (± 2.3) |
| Lamprey CR1 Truncated Seq. | EC50 (nM) | 0.24 | 2.3 | 8.6 | 7.9 | 26 |
| | Fold-Activation (±SEM) | 280 (± 9.7) | 264 (± 2.3) | 267 (± 10.1) | 280 (± 19.4) | 284 (± 13.7) |
| Lamprey CR2 Full Seq. | EC50 (nM) | 0.26 | 1.2 | 1.5 | 4 | 7.4 |
| | Fold-Activation (±SEM) | 88 (±4.1) | 79 (± 2.8) | 93 (± 6.8) | 80 (± 1.7) | 71 (± 5.8) |
| Lamprey CR2 Truncated Seq. | EC50 (nM) | 0.31 | 3.7 | 8.2 | 6.6 | 23.0 |
| | Fold-Activation (±SEM) | 1281 (± 51) | 1232 (± 66) | 1325 (± 16) | 1215 (± 69) | 1183 (± 56) |

Full Seq. = full receptor sequence, Truncated Seq. = Receptor with NTD deleted

DOC = 11-deoxycorticosterone, ALDO = aldosterone

**Corticosteroid-dependent and promoter-dependent activation by corticosteroids of full-length and truncated lamprey CR2.**

Corticosteroids in HEK293 cells with full-length CR2 and the MMTV promoter have slightly higher EC50s and a similar fold-activation compared to full-length CR1 with the MMTV promoter (Table 2). However, similar to our results for lamprey CR1, transcriptional activation of full-length and truncated lamprey CR2 in HEK293 cells is different in the presence of the MMTV and TAT3 promoters (Figure 5, Table 2). In HEK293 cells with TAT3, there is about 30% lower fold-activation for full-length CR2 compared to fold-activation for full-length CR2 in cells with MMTV. In contrast, fold-activation for truncated CR2 is about 15-fold higher in the presence of TAT3 than full-length CR2 and 55 to 60 higher activation than for truncated CR2 with MMTV.

**Corticosteroid-dependent and promoter-dependent activation by corticosteroids of full-length and truncated elephant shark MR and GR.**

To gain an insight into corticosteroid activation of the MR and GR early in their evolution, we compared corticosteroid activation of lamprey CR with activation of the MR and GR in elephant shark (*Callorhinchus milii*), a cartilaginous fish belonging to the oldest group of jawed vertebrates. Elephant shark occupy a key position spanning an ancestral node from which ray-finned fish and terrestrial vertebrates diverged about 450 million years ago from bony vertebrates (58,59).

We previously studied corticosteroid activation of elephant shark GR (12,39) and MR (12) in HEK293 cells containing the MMTV promoter. To complete our dataset for an evolutionary analysis of the CR with elephant shark MR and GR, we investigated corticosteroid activation of elephant shark GR and MR in HEK293 cells containing TAT3. Figure 6 and Table 3 show our results. For full-length elephant shark GR, only corticosterone retains an EC50 value close to the EC50 of lamprey CR1. The other corticosteroids have substantially higher EC50s, and are unlikely to be physiological ligands for this GR. Notably, 11-deoxycortisol has little activity for elephant shark GR (EC50=289 nM), which compares to an EC50 of 1.5 nM for lamprey CR1 and 2.7 nM for lamprey CR2. However, fold-activation of full-length elephant shark GR is higher for than for full-length lamprey CR1 with over 600-fold activation for aldosterone, corticosterone and cortisol. Truncated elephant shark GR loses substantial activity for all corticosteroids indicating that the NTD is important in the EC50 and fold-activation of elephant GR.

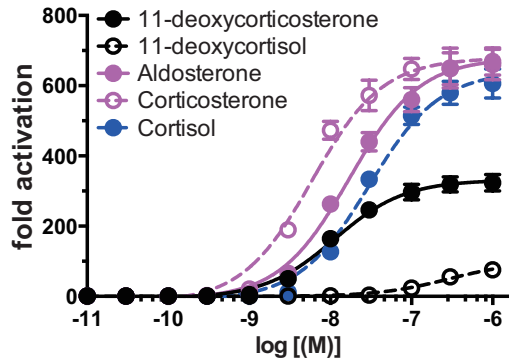
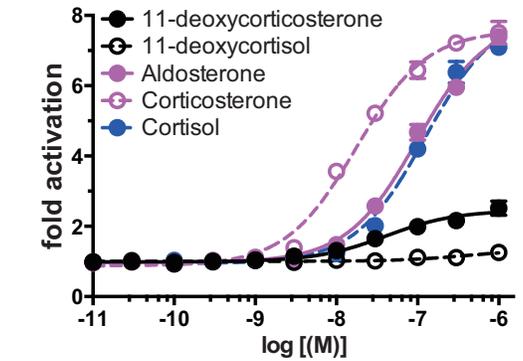
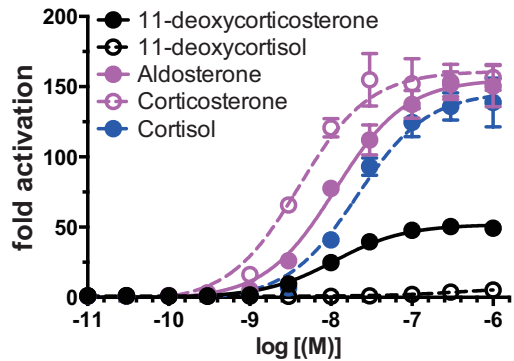
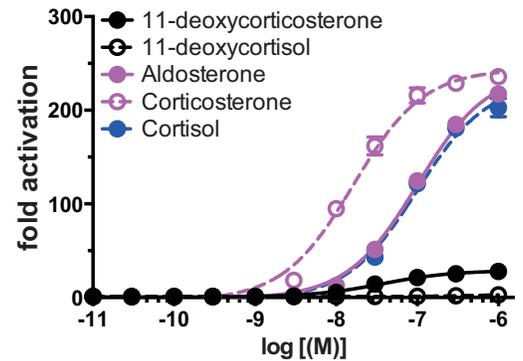
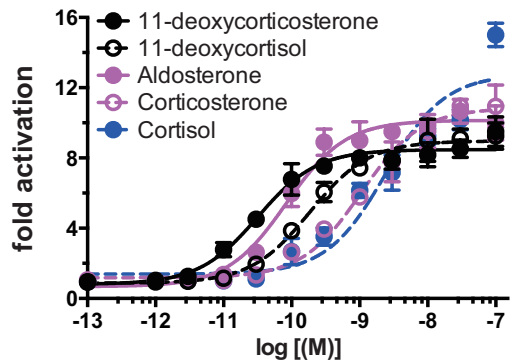
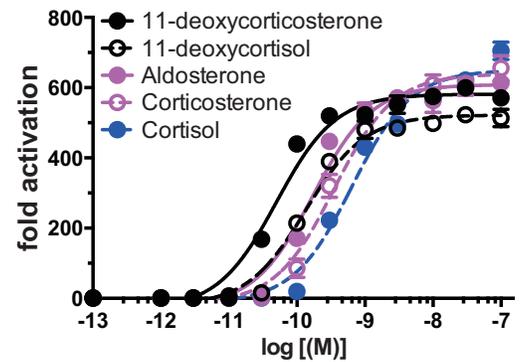

**Fig. 6. Concentration-dependent transcriptional activation by corticosteroids of full length and truncated elephant shark GR and elephant shark MR.** Plasmids for full-length or truncated elephant shark GR were expressed in HEK293 cells with either an MMTV-luciferase promoter or a TAT3-luciferase promoter. Plasmids for full-length or truncated elephant shark MR were expressed in HEK293 cells with a TAT3-luciferase promoter. Cells were treated with increasing concentrations of either aldosterone, cortisol, corticosterone, 11-deoxycortisol, 11-deoxycorticosterone or vehicle alone (DMSO). Results are expressed as means ± SEM, n=3. Y-axis indicates fold-activation compared to the activity of vector treated with vehicle (DMSO) alone as 1. A. Full-length elephant shark GR with MMTV-luc. B. Truncated elephant shark GR with MMTV. C. Full-length elephant shark GR with TAT3. D. Truncated elephant shark GR with TAT3. E. Full-length elephant shark MR with TAT3-luc. F. Truncated elephant shark MR with TAT3.

**Table 3. Corticosteroid Activation of elephant shark GR and MR in HEK293 cells with an MMTV promoter or a TAT3 promoter.**

| MMTV-luc | | DOC | 11-deoxycortisol | ALDO | Corticosterone | Cortisol |
| --- | --- | --- | --- | --- | --- | --- |
| Elephant Shark GR Full Seq. | EC50 (nM) | 11.2 | 289 | 17.1 | 5.9 | 30 |
| | Fold-Activation (±SEM) | 297 (± 22) | 23.4 (± 0.5) | 561 (± 34) | 648 (± 30) | 514 (± 25) |
| Elephant Shark GR Truncated Seq. | EC50 (nM) | 40.5 | - | 98.4 | 17.5 | 124 |
| | Fold-Activation (±SEM) | 2.0 (± 0.1) | 1.1 (± 0.03) | 4.7 (± 0.2) | 6.4 (± 0.23) | 4.2 (± 0.06) |
| Elephant Shark MR Full Seq. * | EC50 (nM) | 0.1 | 0.22 | 0.14 | 0.61 | 1.6 |
| | Fold-Activation (±SEM) | 8.6 (± 0.4) | 8.0 (± 0.5) | 9.6 (± 1.0) | 9.1 (± 0.7) | 10.1 (± 0.9) |
| Elephant Shark MR Truncated Seq. * | EC50 (nM) | 0.09 | 0.28 | 0.26 | 0.58 | 1.06 |
| | Fold-Activation (±SEM) | 6.9 (± 0.15) | 6.6 (± 0.5) | 7.3 (± 0.8) | 8.2 (± 0.4) | 8.1 (± 0.95) |
| **TAT3-luc** | | **DOC** | **11-deoxycortisol** | **ALDO** | **Corticosterone** | **Cortisol** |
| Elephant Shark GR | EC50 (nM) | 11.0 | - | 11.4 | 4.0 | 21 |

| | | | | | | | |
|---|---|---|---|---|---|---|---|
| Full Seq. | Fold-Activation (±SEM) | 48 (± 3.5) | 2.1 (± 0.2) | 137.3 (± 10) | 152 (± 18) | 124 (± 10.2) |
| Elephant Shark GR Truncated Seq. | EC50 (nM) | 37.4 | - | 100.3 | 16.3 | 99.6 |
| | Fold-Activation (±SEM) | 21.4 (± 0.8) | 1.3 (± 0.04) | 125 (± 5) | 216 (± 8) | 122 (± 5.7) |
| Elephant Shark MR Full Seq. | EC50 (nM) | 0.03 | 0.2 | 0.08 | 1.0 | 2.4 |
| | Fold-Activation (±SEM) | 9.5 (± 0.84) | 9.2 (± 0.74) | 10 (± 1) | 11 (± 1.2) | 15 (± 0.7) |
| Elephant Shark MR Truncated Seq. | EC50 (nM) | 0.05 | 0.13 | 0.17 | 0.35 | 0.64 |
| | Fold-Activation (±SEM) | 571 (± 11) | 514 (± 24.6) | 613 (± 17.5) | 656 (± 35.3) | 705 (± 24.6) |

Full Seq. = full receptor sequence, Truncated Seq. = Receptor with NTD deleted

DOC = 11-deoxycorticosterone, ALDO = aldosterone

*: Katsu et al. N-terminal Domain Regulates Steroid Activation of Elephant Shark Glucocorticoid and Mineralocorticoid Receptors. J. Steroid Biochem. Mol .Biol. 210, 105845 (2021)

Unlike full-length elephant shark GR, full-length elephant shark MR in cells with the MMTV promoter has EC50s varying from 0.1 nM (11-deoxycorticosterone) to 1.6 nM (cortisol), which are close to their EC50s for lamprey CR1 and CR2. Truncated elephant shark MR in cells with MMTV retains these low EC50s (Table 3). As found for lamprey CR, in cells with TAT3, removal of the NTD in elephant shark MR leads to a substantial increase of 50-fold in transcriptional activation by corticosteroids (Figure 6, Table 3). These data also support lamprey CR and elephant shark MR as closer to each other than to elephant shark GR.

**Discussion**

Sea lamprey and hagfish, the two extant cyclostomes, occupy a critical position in the evolution of vertebrates (1–3,6), and the sea lamprey CR occupies a critical position in the evolution of the MR and GR in vertebrates (4,8,9,13,38,56), two steroid receptors that are important regulators of vertebrate physiology (10,19,20,22,31,32,36). The assembly of the lamprey germline genome provided an opportunity to determine the response to corticosteroids of full-length CR for comparison to elephant shark GR and MR and provide insights into the evolution of glucocorticoid and mineralocorticoid signaling.  Analysis of the CR in the lamprey genome identified two CRs, CR1 and CR2, which differ only in an insertion of four amino acids in the DBD of CR1.  The DBD in elephant shark MR and GR (12), as well as in other vertebrate MRs and GRs lack these extra four amino acids found in CR1, suggesting that vertebrate MRs and GRs are descended from an ancestral CR2-like gene.  Expression of CR1 and CR2, as determined by RNA-Seq revealed that CR1 comprises over 99% of expressed CR in lamprey kidney, intestine and brain (Table 1).

Our analysis of steroid activation of lamprey CR1 and CR2 found that 11-deoxycorticosterone and 11-deoxycortisol (Figure 1) have the lowest EC50s and a high fold-activation of transcription, consistent with the evidence that these are the circulating steroids in lamprey (26–28).  Overall, full-length lamprey CR1 and CR2 have similar EC50s and fold-activation for corticosteroids in cells transfected with MMTV promoter.  Truncated lamprey CR1 and CR2 also have similar EC50s and fold-activation for corticosteroids in cells transfected with MMTV promoter.  However, in HEK293 cells transfected with TAT3, compared to full-length CR1, full-length CR2 has about 2.25-fold higher activation in the presence of corticosteroids and truncated CR2 has about 4.5-fold higher fold activation than truncated CR1.  It appears that a complex mechanism regulating CR-mediated transcription involving that the NTD and the MMTV and TAT3 promoters evolved before the evolution of distinct GR and MR genes in an ancestral cartilaginous fish.

Analysis of the effect of deleting the NTD on the CR and elephant shark MR and GR on corticosteroid activation of lamprey CR and elephant MR and GR in HEK293 cells with TAT3 support the hypothesis that lamprey CR and elephant shark MR as functionally more similar to one another than to elephant shark GR.  A similar conclusion comes from comparison of the responses of lamprey CRs and elephant shark MR and GR to corticosteroids.

Although 11-deoxycortisol and 11-deoxycorticosterone are the circulating corticosteroids in lamprey, studies of these two steroids in live lamprey find that only 11-deoxycortisol is biologically active (26,27,60).  This contrasts with our results in cell culture in which both steroids, as well as other corticosteroids, stimulate transcription of CR1 and CR2 (Figures 4 and 5).  There may be additional regulatory mechanisms that lead to preferential activity of 11-deoxycortisol in lamprey and inactivity of 11-deoxycorticosterone. It appears that the Atlantic sea lamprey has more secrets to share.

**Materials and Methods**

**Chemical reagents**

Cortisol, corticosterone, 11-deoxycorticosterone, 11-deoxycortisol, and aldosterone were purchased from Sigma-Aldrich. For reporter gene assays, all hormones were dissolved in dimethyl-sulfoxide (DMSO); the final DMSO concentration in the culture medium did not exceed 0.1%.

**Gene expression analysis**

To determine the expression level of CR gene, single-end RNA-seq reads of sea lamprey for seven stages of embryos, intestine and kidney from larval stage, intestine, kidney and liver from parasitic stage, brain, intestine and kidney from adult stage, were downloaded from database of National Center for Biotechnology Information (accession number: PRJNA50489). Reference genome assembly and gene annotation were also downloaded from NCBI database (accession ID: GCF_010993605.1). RNA-seq reads of various tissues were independently aligned to the reference sequences using RSEM (v1.3.3) (57). The relative measure of transcript abundance is FPKM (fragments per kilobase of transcript per million mapped reads).

**Construction of plasmid vectors**

Full-length mineralocorticoid receptor sequence of sea lamprey, *Petromyzon marinus*, was registered in Genbank (accession number: XM_032955475). Based on the registered sea lamprey MR sequence we synthesized DNA containing the full-length sequence. Full-length sea lamprey CR1 was ligated into pcDNA3.1 vector (Invitrogen). CR2 construction was performed using KOD-Plus-mutagenesis kit (TOYOBO). All cloned DNA sequences were verified by sequencing.

**Transactivation assay and statistical methods**

Transfection and reporter assays were carried out in HEK293 cells, as described previously (12). The cells were transfected with 100 ng of receptor gene, reporter gene containing the *Photinus pyralis* luciferase gene and pRL-tk, as an internal control to normalize for variation in transfection efficiency; pRL-tk contains the *Renilla reniformis* luciferase gene with the herpes simplex virus thymidine kinase promoter. All experiments were performed in triplicate. Promoter activity was calculated as firefly (*P. pyralis*)-lucifease activity/sea pansy (*R. reniformis*)-lucifease activity. The values shown are mean ± SEM from three separate experiments, and dose-response data, which were used to calculate the half maximal response (EC50) for each steroid, were analyzed using GraphPad Prism. Comparisons between two groups were performed using paired *t*-test. $P < 0.05$ was considered statistically significant.

**Funding, Contributions and Competing Interests.**

**Funding:** Z.C. gratefully acknowledges the financial support from the China Scholarship Council (CSC Grant No.201808210287). This work was supported by Grants-in-Aid for Scientific Research from the Ministry of Education, Culture, Sports, Science and Technology of Japan (19K067309 to Y.K.), and the Takeda Science Foundation (to Y.K.). M.E.B. was supported by Research fund #3096.

**Author contributions:** Y.K. and M.E.B. designed the research. Y.K. and X.L. carried out the research and analyzed data. R.J., Z.C., Y.K., K.B. performed the cell culture and DNA construction. Y.K. and M.E.B. wrote the paper. All authors gave final approval for publication.

**Competing Interests:** We have no competing interests.

**References**

1. Osório J, Rétaux S. The lamprey in evolutionary studies. Dev Genes Evol. 2008 May;218(5):221-35. doi: 10.1007/s00427-008-0208-1. Epub 2008 Feb 15. PMID: 18274775.

2. Shimeld SM, Donoghue PC. Evolutionary crossroads in developmental biology: cyclostomes (lamprey and hagfish). Development. 2012 Jun;139(12):2091-9. doi: 10.1242/dev.074716. PMID: 22619386.

3. Smith JJ, Kuraku S, Holt C, Sauka-Spengler T, Jiang N, Campbell MS, Yandell MD, Manousaki T, Meyer A, Bloom OE, Morgan JR, Buxbaum JD, Sachidanandam R, Sims C, Garruss AS, Cook M, Krumlauf R, Wiedemann LM, Sower SA, Decatur WA, Hall JA, Amemiya CT, Saha NR, Buckley KM, Rast JP, Das S, Hirano M, McCurley N, Guo P, Rohner N, Tabin CJ, Piccinelli P, Elgar G, Ruffier M, Aken BL, Searle SM, Muffato M, Pignatelli M, Herrero J, Jones M, Brown CT, Chung-Davidson YW, Nanlohy KG, Libants SV, Yeh CY, McCauley DW, Langeland JA, Pancer Z, Fritzsch B, de Jong PJ, Zhu B, Fulton LL, Theising B, Flicek P, Bronner ME, Warren WC, Clifton SW, Wilson RK, Li W. Sequencing of the sea lamprey (Petromyzon marinus) genome provides insights into vertebrate evolution. Nat Genet. 2013 Apr;45(4):415-21, 421e1-2. doi: 10.1038/ng.2568. Epub 2013 Feb 24. PMID: 23435085; PMCID: PMC3709584.

4. Bridgham JT, Carroll SM, Thornton JW. Evolution of hormone-receptor complexity by molecular exploitation. Science. 2006;312(5770):97-101. doi:10.1126/science.1123348.

5. Grillner S, Robertson B. The Basal Ganglia Over 500 Million Years. Curr Biol. 2016 Oct 24;26(20):R1088-R1100. doi: 10.1016/j.cub.2016.06.041. PMID: 27780050.

6. Green SA, Bronner ME. The lamprey: a jawless vertebrate model system for examining origin of the neural crest and other vertebrate traits. Differentiation. 2014 Jan-Feb;87(1-2):44-51. doi: 10.1016/j.diff.2014.02.001. Epub 2014 Feb 20. PMID: 24560767; PMCID: PMC3995830.

7. Ortlund EA, Bridgham JT, Redinbo MR, Thornton JW. Crystal structure of an ancient protein: evolution by conformational epistasis. Science. 2007 Sep 14;317(5844):1544-8. doi: 10.1126/science.1142819. Epub 2007 Aug 16. PMID: 17702911; PMCID: PMC2519897.

8. Bouyoucos IA, Schoen AN, Wahl RC, Anderson WG. Ancient fishes and the functional evolution of the corticosteroid stress response in vertebrates. Comp Biochem Physiol A Mol Integr Physiol. 2021 Oct;260:111024. doi: 10.1016/j.cbpa.2021.111024. Epub 2021 Jul 5. PMID: 34237466.

9. Baker ME, Katsu Y. 30 YEARS OF THE MINERALOCORTICOID RECEPTOR: Evolution of the mineralocorticoid receptor: sequence, structure and function. J Endocrinol. 2017;234(1):T1-T16. doi:10.1530/JOE-16-0661.

10. Bury NR, Sturm A. Evolution of the corticosteroid receptor signalling pathway in fish. Gen Comp Endocrinol. 2007 Aug-Sep;153(1-3):47-56. doi: 10.1016/j.ygcen.2007.03.009. Epub 2007 Mar 24. PMID: 17470371.

11. Carroll SM, Bridgham JT, Thornton JW. Evolution of hormone signaling in elasmobranchs by exploitation of promiscuous receptors. Mol Biol Evol. 2008;25(12):2643-2652. doi:10.1093/molbev/msn204.

12. Katsu Y, Shariful IMD, Lin X, Takagi W, Urushitani H, Kohno S, Hyodo S, Baker ME. N-terminal Domain Regulates Steroid Activation of Elephant Shark Glucocorticoid and Mineralocorticoid Receptors. J Steroid Biochem Mol Biol. 2021 Feb 27:105845. doi: 10.1016/j.jsbmb.2021.105845. Epub ahead of print. PMID: 33652098.

13. Rossier BC, Baker ME, Studer RA. Epithelial sodium transport and its control by aldosterone: the story of our internal environment revisited. Physiol Rev. 2015;95(1):297-340. doi:10.1152/physrev.00011.2014.

14. Evans RM. The steroid and thyroid hormone receptor superfamily. Science. 1988;240(4854):889-895. doi:10.1126/science.3283939.

15. Bridgham JT, Eick GN, Larroux C, et al. Protein evolution by molecular tinkering: diversification of the nuclear receptor superfamily from a ligand-dependent ancestor. PLoS Biol. 2010;8(10):e1000497. Published 2010 Oct 5. doi:10.1371/journal.pbio.1000497.

16. Baker ME. Steroid receptors and vertebrate evolution. Mol Cell Endocrinol. 2019;496:110526. doi:10.1016/j.mce.2019.110526.

17. Whitfield GK, Jurutka PW, Haussler CA, Haussler MR. Steroid hormone receptors: evolution, ligands, and molecular basis of biologic function. J Cell Biochem. 1999;Suppl 32-33:110-22. doi: 10.1002/(sici)1097-4644(1999)75:32+<110::aid-jcb14>3.0.co;2-t. PMID: 10629110.

18. Beato M, Klug J. Steroid hormone receptors: an update. Hum Reprod Update. 2000 May-Jun;6(3):225-36. doi: 10.1093/humupd/6.3.225. PMID: 10874567.

19. Lifton RP, Gharavi AG, Geller DS. Molecular mechanisms of human hypertension. Cell. 2001;104(4):545-556. doi:10.1016/s0092-8674(01)00241-0.

20. Shibata S. 30 YEARS OF THE MINERALOCORTICOID RECEPTOR: Mineralocorticoid receptor and NaCl transport mechanisms in the renal distal nephron. J Endocrinol. 2017;234(1):T35-T47. doi:10.1530/JOE-16-0669.

21. Hanukoglu I, Hanukoglu A. Epithelial sodium channel (ENaC) family: Phylogeny, structure-function, tissue distribution, and associated inherited diseases. Gene. 2016;579(2):95-132. doi:10.1016/j.gene.2015.12.061.

22. Hawkins UA, Gomez-Sanchez EP, Gomez-Sanchez CM, Gomez-Sanchez CE. The ubiquitous mineralocorticoid receptor: clinical implications. Curr Hypertens Rep. 2012;14(6):573-580. doi:10.1007/s11906-012-0297-0.

23. Jaisser F, Farman N. Emerging Roles of the Mineralocorticoid Receptor in Pathology: Toward New Paradigms in Clinical Pharmacology. Pharmacol Rev. 2016;68(1):49-75. doi:10.1124/pr.115.011106.

24. de Kloet ER, Joëls M. Brain mineralocorticoid receptor function in control of salt balance and stress-adaptation. Physiol Behav. 2017 Sep 1;178:13-20. doi: 10.1016/j.physbeh.2016.12.045. Epub 2017 Jan 13. PMID: 28089704.

25. Grossmann C, Almeida-Prieto B, Nolze A, Alvarez de la Rosa D. Structural and molecular determinants of mineralocorticoid receptor signalling. Br J Pharmacol. 2021 Nov 22. doi: 10.1111/bph.15746. Epub ahead of print. PMID: 34811739.

26. Close DA, Yun SS, McCormick SD, Wildbill AJ, Li W. 11-deoxycortisol is a corticosteroid hormone in the lamprey. Proc Natl Acad Sci U S A. 2010 Aug 3;107(31):13942-7. doi: 10.1073/pnas.0914026107. Epub 2010 Jul 19. PMID: 20643930; PMCID: PMC2922276.

27. Shaughnessy CA, Barany A, McCormick SD. 11-Deoxycortisol controls hydromineral balance in the most basal osmoregulating vertebrate, sea lamprey (Petromyzon marinus). Sci Rep. 2020;10(1):12148. Published 2020 Jul 22. doi:10.1038/s41598-020-69061-4.

28. Roberts BW, Didier W, Rai S, Johnson NS, Libants S, Yun SS, Close DA. Regulation of a putative corticosteroid, 17,21-dihydroxypregn-4-ene,3,20-one, in sea lamprey, Petromyzon marinus. Gen Comp Endocrinol. 2014 Jan 15;196:17-25. doi: 10.1016/j.ygcen.2013.11.008. Epub 2013 Nov 25. PMID: 24287339.

29. Baker ME, Funder JW, Kattoula SR. Evolution of hormone selectivity in glucocorticoid and mineralocorticoid receptors [published correction appears in J Steroid Biochem Mol Biol. 2014 Jan;139:104]. J Steroid Biochem Mol Biol. 2013;137:57-70. doi:10.1016/j.jsbmb.2013.07.009.

30. Bury NR. The evolution, structure and function of the ray finned fish (Actinopterygii) glucocorticoid receptors. Gen Comp Endocrinol. 2017 Sep 15;251:4-11. doi: 10.1016/j.ygcen.2016.06.030. Epub 2016 Nov 10. PMID: 27838382.

31. Chrousos GP. Stress and sex versus immunity and inflammation. Sci Signal. 2010 Oct 12;3(143):pe36. doi: 10.1126/scisignal.3143pe36. PMID: 20940425.

32. Grossmann C, Scholz T, Rochel M, Bumke-Vogt C, Oelkers W, Pfeiffer AF, Diederich S, Bahr V. Transactivation via the human glucocorticoid and mineralocorticoid receptor by therapeutically used steroids in CV-1 cells: a comparison of their glucocorticoid and

mineralocorticoid properties. Eur J Endocrinol. 2004 Sep;151(3):397-406. doi: 10.1530/eje.0.1510397. PMID: 15362971.

33. Smith JJ, Antonacci F, Eichler EE, Amemiya CT. Programmed loss of millions of base pairs from a vertebrate genome. Proc Natl Acad Sci U S A. 2009 Jul 7;106(27):11212-7. doi: 10.1073/pnas.0902358106. Epub 2009 Jun 26. PMID: 19561299; PMCID: PMC2708698.

34. Smith JJ, Saha NR, Amemiya CT. Genome biology of the cyclostomes and insights into the evolutionary biology of vertebrate genomes. Integr Comp Biol. 2010 Jul;50(1):130-7. doi: 10.1093/icb/icq023. Epub 2010 Apr 19. PMID: 21558194; PMCID: PMC3140258.

35. Smith JJ, Timoshevskaya N, Ye C, Holt C, Keinath MC, Parker HJ, Cook ME, Hess JE, Narum SR, Lamanna F, Kaessmann H, Timoshevskiy VA, Waterbury CKM, Saraceno C, Wiedemann LM, Robb SMC, Baker C, Eichler EE, Hockman D, Sauka-Spengler T, Yandell M, Krumlauf R, Elgar G, Amemiya CT. The sea lamprey germline genome provides insights into programmed genome rearrangement and vertebrate evolution. Nat Genet. 2018 Feb;50(2):270-277. doi: 10.1038/s41588-017-0036-1. Epub 2018 Jan 22. Erratum in: Nat Genet. 2018 Apr 19;: Erratum in: Nat Genet. 2018 Nov;50(11):1617. PMID: 29358652; PMCID: PMC5805609.

36. Weikum ER, Knuesel MT, Ortlund EA, Yamamoto KR. Glucocorticoid receptor control of transcription: precision and plasticity via allostery. Nat Rev Mol Cell Biol. 2017;18(3):159-174. doi:10.1038/nrm.2016.152.

37. Christopoulos A, Changeux JP, Catterall WA, et al. International Union of Basic and Clinical Pharmacology. XC. multisite pharmacology: recommendations for the nomenclature of receptor allosterism and allosteric ligands. Pharmacol Rev. 2014;66(4):918-947. doi:10.1124/pr.114.008862.

38. Thornton JW. Evolution of vertebrate steroid receptors from an ancestral estrogen receptor by ligand exploitation and serial genome expansions. Proc Natl Acad Sci U S A. 2001 May 8;98(10):5671-6. doi: 10.1073/pnas.091553298. Epub 2001 May 1. PMID: 11331759; PMCID: PMC33271.

39. Katsu Y, Kohno S, Oka K, et al. Transcriptional activation of elephant shark mineralocorticoid receptor by corticosteroids, progesterone, and spironolactone. Sci Signal. 2019;12(584):eaar2668. Published 2019 Jun 4. doi:10.1126/scisignal.aar2668, (n.d.).

40. Gross KL, Cidlowski JA. Tissue-specific glucocorticoid action: a family affair. Trends Endocrinol Metab. 2008;19(9):331-339. doi:10.1016/j.tem.2008.07.009.

41. Rupprecht R, Arriza JL, Spengler D, et al. Transactivation and synergistic properties of the mineralocorticoid receptor: relationship to the glucocorticoid receptor. Mol Endocrinol. 1993;7(4):597-603. doi:10.1210/mend.7.4.8388999.

42. Lu NZ, Cidlowski JA. Glucocorticoid receptor isoforms generate transcription specificity. Trends Cell Biol. 2006 Jun;16(6):301-7. doi: 10.1016/j.tcb.2006.04.005. Epub 2006 May 11. PMID: 16697199.

43. Changeux JP, Christopoulos A. Allosteric Modulation as a Unifying Mechanism for Receptor Function and Regulation. Cell. 2016 Aug 25;166(5):1084-1102. doi: 10.1016/j.cell.2016.08.015. PMID: 27565340.

44. Hollenberg SM, Giguere V, Segui P, Evans RM. Colocalization of DNA-binding and transcriptional activation functions in the human glucocorticoid receptor. Cell. 1987;49(1):39-46. doi:10.1016/0092-8674(87)90753-7.

45. Dahlman-Wright K, Almlöf T, McEwan IJ, Gustafsson JA, Wright AP. Delineation of a small region within the major transactivation domain of the human glucocorticoid receptor that mediates transactivation of gene expression. Proc Natl Acad Sci U S A. 1994;91(5):1619-1623. doi:10.1073/pnas.91.5.1619.

46. Li J, White JT, Saavedra H, et al. Genetically tunable frustration controls allostery in an intrinsically disordered transcription factor [published correction appears in Elife. 2018 Feb 12;7:]. Elife. 2017;6:e30688. Published 2017 Oct 12. doi:10.7554/eLife.30688.

47. Sturm A, Bron JE, Green DM, Bury NR. Mapping of AF1 transactivation domains in duplicated rainbow trout glucocorticoid receptors. J Mol Endocrinol. 2010;45(6):391-404. doi:10.1677/JME-09-0152.

48. Eliezer D. Proteins acting out of (dis)order. Elife. 2017;6:e32762. Published 2017 Nov 21. doi:10.7554/eLife.32762.

49. Fuse H, Kitagawa H, Kato S. Characterization of transactivational property and coactivator mediation of rat mineralocorticoid receptor activation function-1 (AF-1). Mol Endocrinol. 2000 Jun;14(6):889-99. doi: 10.1210/mend.14.6.0467. PMID: 10847590.

50. Fischer K, Kelly SM, Watt K, Price NC, McEwan IJ. Conformation of the Mineralocorticoid Receptor N-Terminal Domain: Evidence for Induced and Stable Structure. Mol Endocrinol. 2010 Oct;24(10):1935-48. Doi: 10.1210/Me.2010-0005. Epub 2010 Aug 4. PMID: 20685853; PMCID: PMC5417395.

51. Pascual-Le Tallec L, Lombès M. The mineralocorticoid receptor: a journey exploring its diversity and specificity of action. Mol Endocrinol. 2005 Sep;19(9):2211-21. doi: 10.1210/me.2005-0089. Epub 2005 Mar 31. PMID: 15802372.

52. Cato AC, Skroch P, Weinmann J, Butkeraitis P, Ponta H. DNA sequences outside the receptor-binding sites differently modulate the responsiveness of the mouse mammary tumour virus promoter to various steroid hormones. EMBO J. 1988 May;7(5):1403-10. PMID: 2842149; PMCID: PMC458390.

53. Beato M, Arnemann J, Chalepakis G, Slater E, Willmann T. Gene regulation by steroid hormones. J Steroid Biochem. 1987;27(1-3):9-14. doi: 10.1016/0022-4731(87)90288-3. PMID: 2826895.

54. Iñiguez-Lluhí JA, Pearce D. A common motif within the negative regulatory regions of multiple factors inhibits their transcriptional synergy. Mol Cell Biol. 2000 Aug;20(16):6040-50. doi: 10.1128/mcb.20.16.6040-6050.2000. PMID: 10913186; PMCID: PMC86080.

55. Katsu Y, Oana S, Lin X, Hyodo S, Baker ME. Aldosterone and dexamethasone activate African lungfish mineralocorticoid receptor: Increased activation after removal of the amino-terminal domain. J Steroid Biochem Mol Biol. 2022 Jan;215:106024. doi: 10.1016/j.jsbmb.2021.106024. Epub 2021 Nov 10. PMID: 34774724.

56. Kassahn KS, Ragan MA, Funder JW. Mineralocorticoid receptors: evolutionary and pathophysiological considerations. Endocrinology. 2011 May;152(5):1883-90. doi: 10.1210/en.2010-1444. Epub 2011 Feb 22. PMID: 21343255.

57. Li B, Dewey CN. RSEM: accurate transcript quantification from RNA-Seq data with or without a reference genome. BMC Bioinformatics. 2011 Aug 4;12:323. doi: 10.1186/1471-2105-12-323. PMID: 21816040; PMCID: PMC3163565.

58. Venkatesh B, Lee AP, Ravi V, Maurya AK, Lian MM, Swann JB, Ohta Y, Flajnik MF, Sutoh Y, Kasahara M, Hoon S, Gangu V, Roy SW, Irimia M, Korzh V, Kondrychyn I, Lim ZW, Tay BH, Tohari S, Kong KW, Ho S, Lorente-Galdos B, Quilez J, Marques-Bonet T, Raney BJ, Ingham PW, Tay A, Hillier LW, Minx P, Boehm T, Wilson RK, Brenner S, Warren WC. Elephant shark genome provides unique insights into gnathostome evolution. Nature. 2014 Jan 9;505(7482):174-9. doi: 10.1038/nature12826. Erratum in: Nature. 2014 Sep 25;513(7519):574. Erratum in: Nature. 2020 Dec;588(7837):E15. PMID: 24402279; PMCID: PMC3964593.

59. Hara Y, Yamaguchi K, Onimaru K, Kadota M, Koyanagi M, Keeley SD, Tatsumi K, Tanaka K, Motone F, Kageyama Y, Nozu R, Adachi N, Nishimura O, Nakagawa R, Tanegashima C, Kiyatake I, Matsumoto R, Murakumo K, Nishida K, Terakita A, Kuratani S, Sato K, Hyodo S, Kuraku S. Shark genomes provide insights into elasmobranch evolution and the origin of vertebrates. Nat Ecol Evol. 2018 Nov;2(11):1761-1771. doi: 10.1038/s41559-018-0673-5. Epub 2018 Oct 8. PMID: 30297745.

60. Shaughnessy CA, McCormick SD. Functional characterization and osmoregulatory role of the Na+-K+-2Cl- cotransporter in the gill of sea lamprey (Petromyzon marinus), a basal vertebrate. Am J Physiol Regul Integr Comp Physiol. 2020;318(1):R17-R29. doi:10.1152/ajpregu.00125.2019.